\documentclass[reprint,superscriptaddress]{revtex4-1}
\usepackage[utf8]{inputenc}

\usepackage{amsmath}
\usepackage{amssymb}
\usepackage{bm}
\usepackage{graphicx}
\usepackage{xcolor}
\usepackage{hyperref}
\hypersetup{
	colorlinks = true,
	linkcolor  = blue,
	citecolor  = blue,
	filecolor  = magenta,
	urlcolor   = cyan
}
\usepackage{ulem}
\usepackage{enumitem}
\usepackage{textgreek}

\newcommand{\p}{\partial}
\newcommand{\e}{\bm{\hat{e}}}

\newcommand{\magn}{\bm{m}}
\newcommand{\Energy}{E}

\newcommand{\heff}{\bm{h_{\rm eff}}}

\newcommand{\Exchange}{A}
\newcommand{\DM}{D}
\newcommand{\Anisotropy}{K}
\newcommand{\dm}{\lambda}

\newcommand{\storque}{\beta}

\newcommand{\ldw}{\ell_{\rm w}}

\newcommand{\Skyrmion}{Q}
\newcommand{\skyrmion}{q}
\newcommand{\winding}{\kappa}

\begin{document}

\title{Meron configurations in easy-plane chiral magnets}
\author{David Bachmann}
\affiliation{Physikalisches Institut, Albert-Ludwigs-Universit\"at Freiburg, Hermann-Herder-Str.\ 3,
D-79104 Freiburg, Germany}
\author{Michail Lianeris}
\affiliation{Department of Mathematics and Applied Mathematics, University of Crete, 70013 Heraklion, Crete, Greece}
\affiliation{Foundation for Research and Technology (FORTH), 70013 Heraklion, Crete, Greece}
\affiliation{Department of Electrical and Information Engineering, Polytechnic University of Bari , 70126 Bari BA, Italy}
\author{Stavros Komineas}
\affiliation{Department of Mathematics and Applied Mathematics, University of Crete, 70013 Heraklion, Crete, Greece}
\affiliation{Foundation for Research and Technology (FORTH), 70013 Heraklion, Crete, Greece}
\date{\today}

\begin{abstract}
We demonstrate the existence and study in detail the features of chiral bimerons which are static solutions in an easy-plane magnet with the Dzyaloshinskii-Moriya (DM) interaction.
These are skyrmionic textures with an integer topological charge and they present essential analogies to the meron configurations introduced in the context of quark confinement in the O(3) nonlinear $\sigma$-model.
We employ a M\"obius transformation to show that, for weak chirality, bimeron configurations approach Belavin-Polyakov (BP) solutions characterized by tightly bound vortex and antivortex parts of the same size.
Stronger chirality induces different vortex and antivortex sizes and also a detachment of merons, suggesting the possibility for a topological phase transition.
Exploiting the fact that bimerons of opposite topological charges may exist in the same material, we demonstrate numerically a mechanism to generate meron pairs.
\end{abstract}

\maketitle

\section{Introduction}
\label{sec:introduction}

Merons are localized configurations that possess one-half topological charge and they are relevant in theories ranging from high-energy physics to condensed matter.
Their name reflects the fact that a meron can be considered as a part (greek: \textmu\textepsilon\textrho\textomikron\textvarsigma) of a soliton with an integer topological charge \cite{1977_PLB_Gross,1978_NPB_Gross}.
For large distances, or large couplings, the formation of merons is favored due to their logarithmically divergent energy, and they are considered to offer a possible mechanism leading to quark confinement.

In magnetic films with the chiral {\it Dzyaloshinskii-Moriya} (DM) interaction, topological solitons with integer-valued topological charge are the magnetic {\it skyrmions} \cite{1994_JMMM_BogdanovHubert,EverschorMasellReeveKlaeui_JAP2018, 2022_JPD_BoHu}.
Merons were introduced in the context of the nonlinear O(3) $\sigma$-model, a prototype model corresponding to the time-independent {\it Landau-Lifshitz} equation, in Ref.~ \cite{1978_NPB_Gross}.
Under {\it M\"obius transformations}, an axially symmetric skyrmion of topological charge one is decomposed into two spatially separated merons that may have different sizes.

We study chiral ferromagnets with {\it easy-plane anisotropy} which support skyrmionic configurations consisting of two merons, known as {\it bimerons} \cite{2019_PRAppl_MoonHwang,2019_NatComm_GaoYeIm,2019_PRB_GoebelTretiakov,2020_PRB_XuFengBellaiche,2020_NPJ-CM_LiZhou,2021_PhysRep_BoergeMertigTretiakov}.
The two constituent parts are a vortex and an antivortex of different polarities, each contributing one-half of the topological charge.
While a single vortex may be energetically favored by the DM interaction, it is a challenging problem whether a composite configuration including vortices of both windings, i.e., a vortex and an antivortex, yields a stable configuration.

Bimeron structures have been studied in non-chiral magnets \cite{2009_PRB_Kim,2017_PRL_KharkovMostovoy,2021_APL_HanLeeIm} and they have been observed in confined geometries \cite{2019_NatComm_GaoYeIm,2021_APL_HanLeeIm} as well as in antiferromagnetic \textalpha-${\rm Fe}_2{\rm O}_3$ films \cite{2021_Nature_HariomRadaelli}.
Observations of square lattices of chiral merons \cite{2009_PRB_YiHan} have been reported in Ref.~\cite{2018_nature_YuTokura} and their stabilization was investigated within a Ginzburg–Landau model \cite{2020_JNS_LiMelcher}.

The chiral bimerons presented here are directly related to the meron configurations constructed in \cite{1978_NPB_Gross}.
First, the configuration is asymmetric (the two merons have different sizes) and a M\"obius transformation gives a skyrmion including two scales.
Second, tuning the chirality parameter allows detaching of the constituent parts.
A further remarkable feature is that the far field of the chiral bimeron is algebraic, similar to the O(3) $\sigma$ model, despite the presence of anisotropy which typically induces exponential decay.

We exploit the possible coexistence of oppositely-charged bimerons in DM magnets and we numerically demonstrate a remarkable process for a smooth generation of a bimeron.
A straightforward iteration of this mechanism can yield a proliferation scheme for bimerons that overcomes topological constraints, opening the possibility for a topological phase transition such as Berezinskii-Kosterlitz-Thouless (BKT).

The paper is organized as follows.
In Sec.~\ref{sec:chiralModel}, the model for easy-plane chiral ferromagnets is presented.
In Sec.~\ref{sec:bimeron}, the numerical solutions for bimerons are given.
In Sec.~\ref{sec:profile}, details of the bimeron profile are discussed, and the relation to the O(3) merons is quantified.
In Sec.~\ref{sec:unbinding}, a mechanism for the generation of bimerons is discussed.
Sec.~\ref{sec:conclusions} contains our concluding remarks.

\section{Easy-plane chiral magnet}
\label{sec:chiralModel}

We consider a ferromagnetic film with easy-plane anisotropy and the Dzyaloshinskii-Moriya interaction.
The magnetic energy is
\begin{equation}  \label{eq:energy0}
\begin{split}
\Energy & = \Exchange \int \p_\mu \magn\cdot\p_\mu\magn\, d^2x + \Anisotropy \int m_3^2\, d^2x  \\
& + \DM \int \e_\mu\cdot (\p_\mu\magn\times\magn)\, d^2x
\end{split}
\end{equation}
where $\magn = (m_1,m_2,m_3)$ is the normalized magnetization vector, $\mu=1,2$, $\e_\mu$ denote the unit vectors in the respective directions, $\Exchange$ is the symmetric exchange parameter, $\Anisotropy$ is the anisotropy parameter, and $\DM$ is the DM or antisymmetric exchange parameter.

The dynamics of the magnetization vector is described by the Landau-Lifshitz equation as obtained from the energy functional \eqref{eq:energy0}.
Using $\ldw=\sqrt{\Exchange/\Anisotropy}$ as the unit of length, we obtain the dimensionless form
\begin{equation} \label{eq:LLG}
\p_\tau\magn = -\magn \times \heff + \alpha \magn\times\p_\tau\magn
\end{equation}
where we include Gilbert damping with parameter $\alpha$, and the effective field reads
\begin{equation}  \label{eq:effectiveField}
\heff = \Delta\magn + m_3 \e_3 - 2\dm \left( \e_\mu\times\p_\mu\magn \right)
\end{equation}
and includes the dimensionless DM parameter
\begin{equation}  \label{eq:parameterDMI}
\dm = \frac{\DM}{2\sqrt{\Exchange\Anisotropy}}.
\end{equation}

\begin{figure}[t]
    \includegraphics[width=8.0cm]{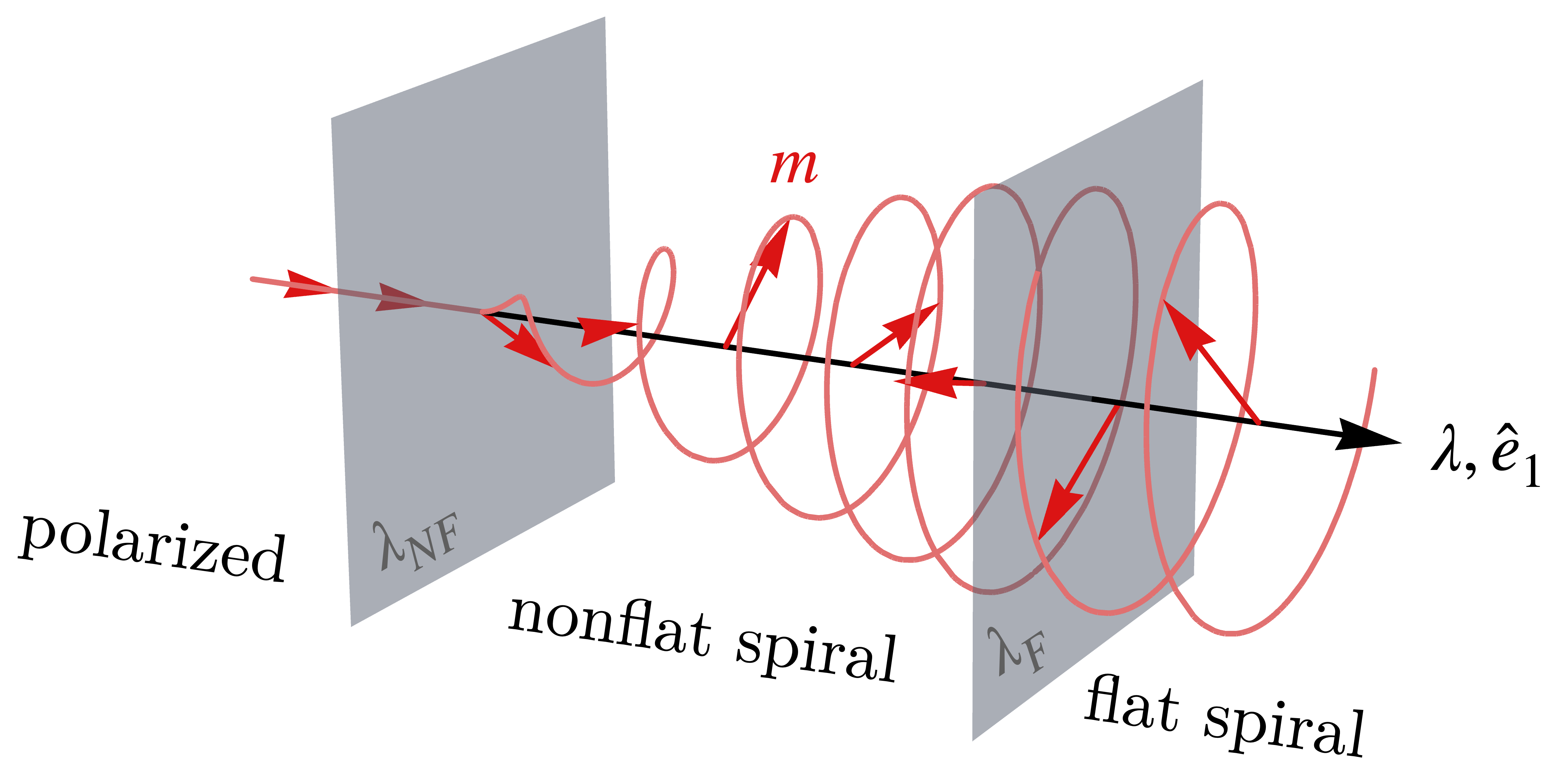}
    \caption{Ground states and phase transitions for an easy-plane chiral ferromagnet vs model parameter $\dm$.
    We have the polarized state for $\dm < \dm_{NF}$, a nonflat spiral for $\dm_{NF} < \dm < \dm_F$ and a flat spiral for $\dm > \dm_F$.
    }
    \label{fig:phases}
\end{figure}

This model appears to be similar to that for easy-axis ferromagnets where a spiral is the ground state for strong enough DM interaction \cite{1994_JMMM_BogdanovHubert}.
In this spiral solution, the magnetization vector rotates in the plane perpendicular to the direction in which the magnetization varies, hence we may call this a {\it flat} spiral.
Despite the apparent similarity of the models, the present case of easy-plane anisotropy allows for an additional phase \cite{2012_PRB_MuelbauerZheludev}, where the phase transitions occur at two critical values of the DM parameter \cite{2002_PRB_ChovanPapanicolaou}
\begin{equation} \label{eq:dm1_dm2}
\dm_{NF} = \frac{1}{2},\qquad
\dm_F \approx 0.705.
\end{equation}
As illustrated in Fig.~\ref{fig:phases},
for weak DM interaction, $\dm < \dm_{NF}$, the fully {\it polarized} state is the ground state with the magnetization vector aligning with an easy plane direction (without loss of generality, we may assume $\magn=\e_1$).
By increasing $\dm$, we enter an intermediate phase in the form of a {\it nonflat} spiral at $\dm=\dm_{NF}$.
The spiral presents a rotation of the projection of $\magn$ on the $(23)$ plane as we move along the $x$ axis and, at the same time, the component $m_1$ oscillates around a nonzero value.
The period of the spiral tends to infinity for $\dm\to\dm_{NF}$ (from above) while the component $m_1$ approaches unity in the same limit.
As $\dm$ increases above $\dm_{NF}$, $m_1$ decreases and it vanishes at $\dm=\dm_F$ where the flat spiral is obtained with $\magn$ perpendicular to $\e_1$ and rotating in the $(23)$ plane.
For $\dm > \dm_F$, the flat spiral is the ground state and its period decreases with increasing $\dm$.

\section{Bimeron solutions}
\label{sec:bimeron}

In a magnetic film (a two-dimensional system), skyrmionic textures \cite{2014_PRX_SumilanRanderia,2016_PRB_RowlandRanderia,2015_PRB_LinSaxenaBatista,2016_APL_VousdenFangohr} and vortices are excited states above the polarized state in the regime $0 < \dm < \dm_{NF}$.
For vortices, the winding of the in-plane magnetization vector, as we rotate around the vortex center, may follow the same or the opposite sense of rotation.
We define accordingly the {\it winding number}, or {\it vortex number}, $\winding=\pm 1$.
We call {\it vortices} those with a positive winding number and {\it antivortices} those with a negative winding number.
The sign of the out-of-plane component of the magnetization in the central region of the vortex (vortex core) defines the vortex {\it polarity}.
For a vortex with positive winding, the orientation of the in-plane magnetization component with respect to the radial direction gives the {\it helicity}.

In chiral magnets, certain vortex configurations are energetically favored by the DM interaction as this gives a negative contribution for particular swirling magnetic configurations. 
This is analogous to the effect of the DM interaction for skyrmions.
This means that a vortex (or a skyrmion) with only one of the two possible windings can be an energy minimum.
Specifically, for the energy \eqref{eq:energy0}, vortices are favored for positive polarity and helicity $-\pi/2$, or negative polarity and helicity $\pi/2$.
Regarding the vortex profile, it is an unusual fact that the magnetization field for a chiral vortex, decays following a power law, as shown by standard asymptotic analysis \cite{2002_PRB_ChovanPapanicolaou}.
This is due to the DM interaction and despite the presence of anisotropy that typically gives exponential decay for vortex configurations.
No isolated static antivortex solutions are found within model \eqref{eq:energy0}.

Magnetic configurations are characterized by the {\it skyrmion number} defined as
\begin{equation}  \label{eq:skyrmionNumber}
\Skyrmion = \frac{1}{4\pi} \int \skyrmion\, d^2x,\quad 
\skyrmion = \magn\cdot(\p_1\magn\times \p_2\magn)
\end{equation}
where $\skyrmion$ is a {\it topological density} and it plays the role of the local vorticity.
Vortices have $\Skyrmion = \pm 1/2$, where the sign depends on their winding number and polarity.
The question arises whether solutions that represent stable skyrmions, i.e., topological solitons with integer skyrmion number, are possible in the easy-plane case.

A skyrmion with $\Skyrmion = \pm 1$ may be constructed from a vortex paired with an antivortex of opposite polarities.
Using the {\it stereographic projection} of the magnetization vector
\begin{equation} \label{eq:Omega}
\Omega = \frac{m_1+i m_2}{1+m_3}.
\end{equation}
such configurations are given by the rational map
\begin{equation}  \label{eq:BPmerons}
\Omega_\mu = \frac{z-ia_1}{z-ia_2},
\end{equation}
where $a_1, a_2 \in \mathbb{R}$, and $z=x+iy$ is the position variable on the complex plane, gives a vortex centered at position $(x,y)=(0,a_1)$ and an antivortex centered at $(0,a_2)$.
Since the absolute position of the resulting two-vortex configuration can be shifted by simple spatial translations, only their separation, given by the difference $a_1-a_2$, is significant.
The vortex exhibits a positive out-of-plane magnetization, i.e., $m_3>0$, whereas the antivortex has $m_3 < 0$.

Within the O(3) nonlinear $\sigma$-model, configurations obtained by rational maps such as \eqref{eq:BPmerons} are exact solutions.
In particular, solutions described by Eq.~\eqref{eq:BPmerons} necessarily feature symmetric vortex and antivortex configurations having the same size.
The two vortices are considered as merons, each occupying one-half of the plane.
Consequently, we can assign to each meron a radius (assume $a_1>a_2$)
\begin{equation} \label{eq:vortexRadius}
R = \frac{a_1-a_2}{2}.
\end{equation}

\begin{figure}[t]
    \centering
    (a) \includegraphics[width=8.0cm]{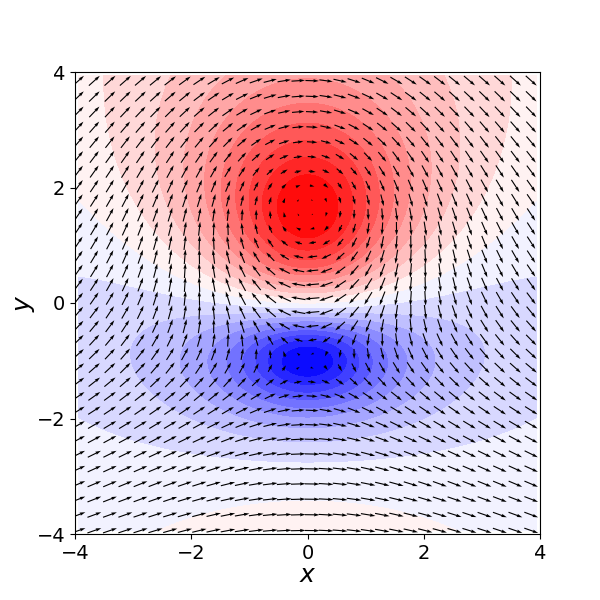}
    (b) \includegraphics[width=8.0cm]{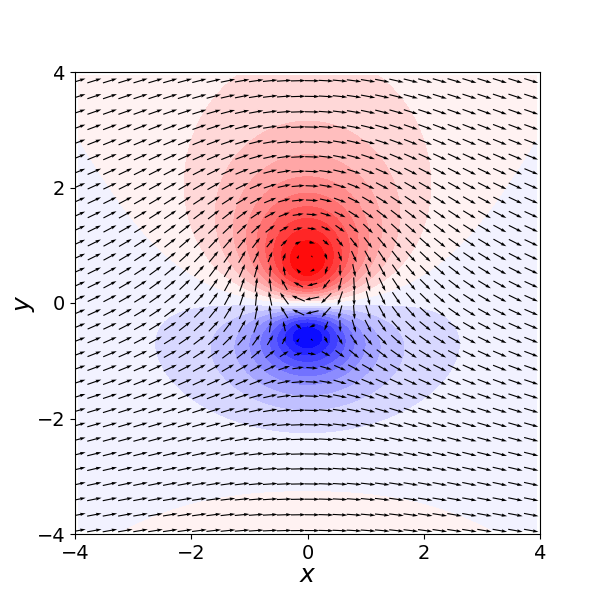}
    \caption{
    Static bimeron solution of model \eqref{eq:LLG}.
    Contour plots for $m_3$ are colored, where red indicates $m_3 > 0$ and blue indicates $m_3 < 0$.
    The center of the bimeron, defined to be at the point where $m_1=-1$, has been placed at the origin.
    The skyrmion number is $\Skyrmion = 1$.
    (a) A bimeron for parameter value $\dm=0.4$.
    The centers of the vortex and antivortex, defined to be at the point where $m_3=\pm 1$, are located on the $y$ axis, at $y=1.66$ and $y=-1.01$, respectively, in units of $\ldw$.
    (b) A bimeron for $\dm=0.32$.
    The vortex center is at $y=0.71$ and the antivortex at $y=-0.56$.
    }
    \label{fig:bimeron}
\end{figure}

In order to find static bimerons, we apply an energy minimization algorithm.
This is equivalent to simulating Eq.~\eqref{eq:LLG} with maximum damping. 
Due to the long range of bimeron configurations, we employ {\it stretched coordinates} $\xi,\eta$, where $x = \tanh\xi,\, y = \tanh\eta$ with $-\pi/2 < \xi,\eta < \pi/2$,
resulting in a lattice in $x,y$ with non-uniform spacing that effectively extends to infinity in all directions.
We typically use a $400\times400$ square lattice with a minimum spacing of $0.08$ (in dimensionless units) at the origin.

We use as an initial condition the form \eqref{eq:BPmerons}.
The numerical relaxation results in {\it asymmetric} meron pairs, i.e., two merons of different sizes.
Fig.~\ref{fig:bimeron} shows two example configurations obtained for two different values of the parameter $\dm$.
The vortex has an almost axially symmetric profile around its center while the antivortex is elongated.
The elongation is more pronounced for larger values of $\dm$, as in Fig.~\ref{fig:bimeron}a, while the antivortex profile is getting closer to an axially symmetric one for lower values of $\dm$, as in Fig.~\ref{fig:bimeron}b.

The antivortex elongation has been noted in Ref.~\cite{2019_PRAppl_MoonHwang} and it is apparent in numerical results showing vortex collections in Ref.~\cite{2015_PRB_LinSaxenaBatista}.
In Ref.~\cite{2021_QTS_Schroers} an elongated vortex is found as an exact solution in a specific solvable $\sigma$ model with DM interaction and easy-plane anisotropy.
The apparent similarity is promising in order to explain the present numerical results but the connection of model \eqref{eq:LLG} with the solvable $\sigma$ model is not straightforward. 

We find that the bimeron solutions exhibit an algebraically (power law) decaying far field with $|\Omega| \sim 1/r^2$.
This feature is shared with the merons in Eq.~\eqref{eq:BPmerons} for the O(3) model, which actually give $|\Omega| \sim 1/r$.
A corresponding result is found by standard asymptotic analysis that gives an algebraic behavior for a single chiral vortex configuration \cite{2002_PRB_ChovanPapanicolaou}.
The power law for a single vortex or a bimeron configuration is an unusual behavior as the presence of anisotropy typically induces exponential decay in vortex configurations.

While $m_3 < 0$ at the antivortex core region, $m_3$ becomes positive in a region well below the antivortex.
The domain with $ m_3>0$ can be discerned in Fig.~\ref{fig:bimeron} by the change in the color well below the antivortex core region.
For example, for $\dm=0.4$, we have $m_3>0$ below $y=-3.43$ on the $y$ axis.
This feature represents a difference between the chiral bimeron configuration and the configuration \eqref{eq:BPmerons} at large distances.

\begin{figure}[t]
    \centering
    \includegraphics[width=8.0cm]{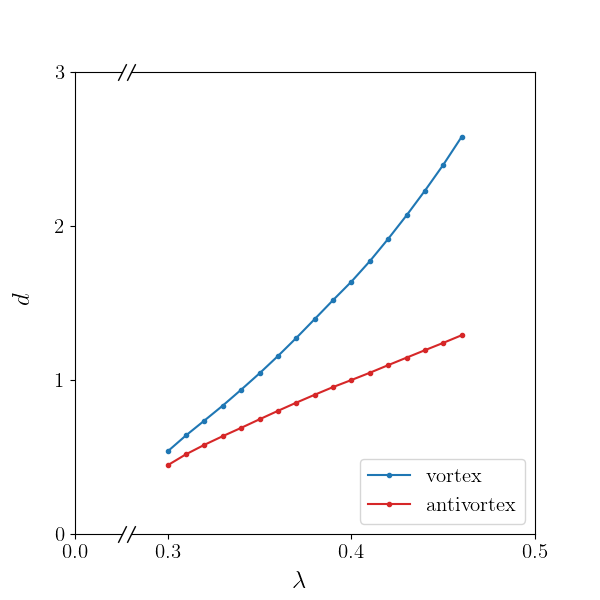}
    \caption{
    Distances of the vortex (upper, blue line) and the antivortex (lower, red line) centers from the bimeron center as a function of the parameter $\dm$.
    Points present numerical results, connected by solid lines.
    For small $\dm$, the vortex and the antivortex are progressively placed more symmetrically on either side of the bimeron center.
    For $\dm\ge0.5$, at the point of transition to the non-flat spiral, the vortex is expected to get completely detached.
    }
    \label{fig:bimeron_size}
\end{figure}

Bimeron solutions are found for a range of parameter values.
We present results for bimeron configurations down to $\dm = 0.28$.
It is numerically challenging to find bimeron solutions for small $\dm$ as an increasingly finer spatial resolution would be needed.
Fig.~\ref{fig:bimeron_size} shows the distances of the vortex and the antivortex centers (the points where $m_3=\pm1$) from the bimeron center (the point where $m_1=-1$) as $\dm$ is varied.
As the value of $\dm$ decreases, the centers of the vortex and the antivortex approach each other and they move progressively to locations symmetrically placed on opposite sides of the bimeron center.

We expect that bimerons exist down to $\dm \to 0$.
In this limit, the DM and anisotropy energy terms decrease and the exchange term dominates.
Thus, the configuration is expected to approach the rational map \eqref{eq:BPmerons} with a decreasing distance between the merons, that is, $a_1-a_2 \to 0$.
This picture is supported by the results in Fig.~\ref{fig:bimeron_size}.
The issue is discussed further in Sec.~\ref{sec:profile}.

A larger parameter $\dm$ gives a larger separation between the two merons.
Past the phase transition to the nonflat spiral, for $\dm \ge \dm_{NF} = 1/2$, we expect no bimeron solutions.
Specifically, for $\dm \ge \dm_{NF}$, a single vortex has negative energy \cite{2002_PRB_ChovanPapanicolaou} and it is expected to detach completely from the antivortex.
As seen in Fig.~\ref{fig:bimeron_size}, this behavior is supported by the numerical simulations which converge to bimeron configurations with increasing meron separation as $\dm \to \dm_{NF}$.

\begin{figure}[t]
    \centering
    \includegraphics[width=8.0cm]{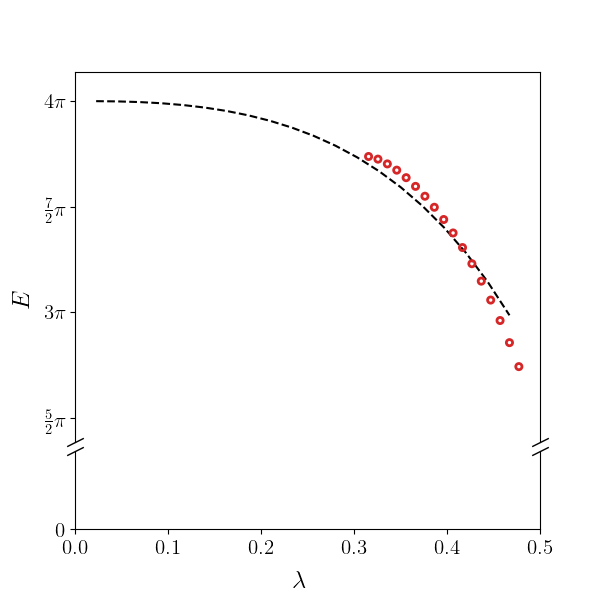}
    \caption{
    The energy of the bimeron as a function of the parameter $\dm$.
    Results from numerical simulations are given by circles.
    The dotted line shows the asymptotic result \eqref{eq:energySmallSkyrmion} for the axially symmetric skyrmions and it is shown for comparison.
    }
    \label{fig:energy}
\end{figure}

Figure~\ref{fig:energy} shows the energy of the bimeron as a function of the parameter $\dm$.
The DM and anisotropy energies decrease in absolute value and they go to zero (from negative and positive values respectively) as $\dm\to 0$, and the bimeron size goes to zero, too.
The exchange energy decreases and it is approaching the value $4\pi$ for $\dm\to0$, which is the value for the bimeron profile \eqref{eq:skyrmion}.
For small $\dm$, the total energy is approaching the value $\Energy=4\pi$ thus supporting the argument that the bimeron configuration is given by \eqref{eq:skyrmion} with $a\to 0$.
A similar situation has been studied for the axially symmetric skyrmion where the asymptotic result for the energy is \cite{2020_NL_KomineasMelcherVenakides,2020_ARMA_BernandMuratovSimon}
\begin{equation} \label{eq:energySmallSkyrmion}
\Energy = 4\pi \left( 1 + \frac{\dm^2}{\ln\dm} \right),\qquad \dm \ll 1.
\end{equation}
In Fig.~\ref{fig:energy}, we tentatively plot formula \eqref{eq:energySmallSkyrmion} and we find that it is in agreement with the present numerical results for small $\dm$.
This can be explained by the fact that the exchange interaction is dominant for $\dm\ll 1$ and the arguments of Refs.~\cite{2020_NL_KomineasMelcherVenakides,2020_ARMA_BernandMuratovSimon} for the asymptotic calculation of the energy can be applied also in the present case.
For larger $\dm$, the energy decreases.
The results in Fig.~\ref{fig:energy} indicate that the energy attains a positive value at $\dm=\dm_{NF}=0.5$.
For $\dm>\dm_{NF}$ the bimeron is not expected to be stable as explained earlier.
A phase transition to a nonflat spiral will occur at that point and the energy would drop discontinuously to negative values.
A full mathematical treatment of these issues would be needed in order to obtain precise results for the phase transitions at $\dm\to 0$ and at $\dm\to\dm_{NF}$, but this is beyond the scope of the present paper.

The fact that the vortex and the antivortex are centered on the $y$ axis for all bimerons presented here, such as in Fig.~\ref{fig:bimeron}, is dictated by the choice of the far field magnetization, $\magn=\e_1$, that is, by the choice for the spontaneously broken symmetry.
If the bimeron configuration would be rotated in space, keeping the far field fixed, this would necessarily have to be accompanied by a change of the helicity of the vortex and thus an increase of the DM energy.
A related fact is that chiral bimerons have been found here as static solutions within the easy-plane magnet, in contrast to vortex-antivortex dipoles in standard models (without the chiral interaction) where the pair is necessarily non-static, i.e., rotating \cite{1985_JETP_PokrovskiiUimin,2007_PRL_Komineas}.
These striking features of chiral bimerons originate in the invariance of DM interaction under simultaneous rotations in real and magnetization space.

We have repeated the simulations including the magnetostatic field and we have verified that the bimeron pair does exist in this case, too.
Therefore, bimerons can be realistically expected to be observed experimentally in a magnetic material with easy-plane anisotropy and chiral interaction.

\section{Chiral Bimeron profile}
\label{sec:profile}

For a quantitative study of the bimeron configuration, we consider a {\it M\"obius transformation} defined by
\begin{equation} \label{eq:Moebius}
\frac{1}{iw} = \frac{z - ia_1}{z - ia_2}
\end{equation}
where
\[
w=u+iv
\]
is the transformed variable in the complex plane.
If this is applied to the configuration \eqref{eq:BPmerons}, we obtain
\begin{equation} \label{eq:skyrmion}
\Omega_S = \frac{1}{iw} = \frac{1}{r}\,e^{-i(\phi+\pi/2)}
\end{equation}
where $(r,\phi)$ are the polar coordinates for $w$.
The result manifestly represents an axially symmetric antiskyrmion with a unit radius.

\begin{figure}[t]
    \centering
   (a) \includegraphics[width=8.0cm]
   {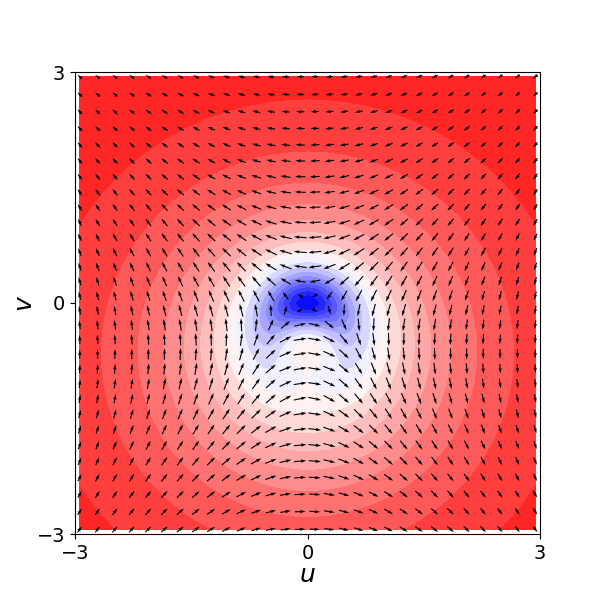}
    (b) \includegraphics[width=8.0cm]{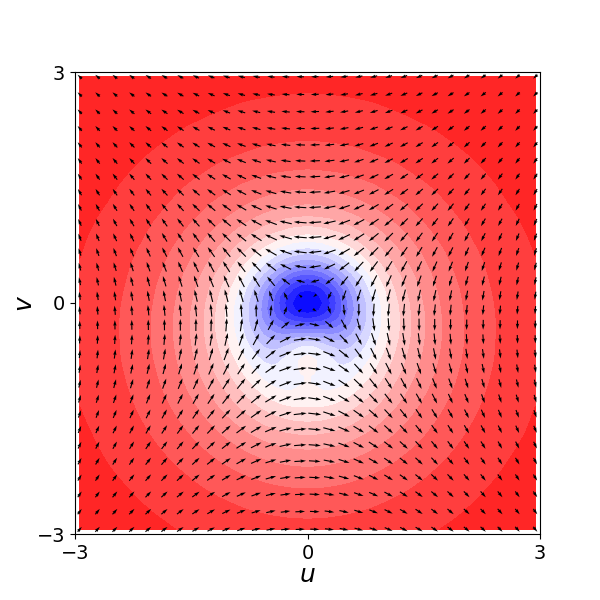}
    \caption{
    (a) Contour plots for the M\"obius transformed configuration \eqref{eq:Omega-tilde} for the bimeron solutions of Fig.~\ref{fig:bimeron} for (a) $\dm=0.4$, and (b) $\dm=0.32$.
    The transformation has produced a skyrmion.
    Blue and red indicate $m_3>0$ and $m_3<0$, respectively.
    The antivortex has been mapped to the center of the skyrmion and the vortex to its periphery.
    The contours in the red region and in the central part of the blue region are approximately circles.
    }
    \label{fig:skyrmion}
\end{figure}

Chiral bimerons share with the configurations considered in Ref.~\cite{1978_NPB_Gross} the salient property of being composed of merons with different radii (cf. Fig.~\ref{fig:bimeron}), unlike the symmetric bimeron configurations in \eqref{eq:BPmerons}.
In order to quantify this, we apply the M\"obius transformation \eqref{eq:Moebius}, with $a_1, a_2$ chosen to coincide with the locations where $m_3=\pm1$, to our bimeron solutions $\Omega(z)$ and obtain
\begin{equation} \label{eq:Omega-tilde}
\tilde{\Omega}(w) = \Omega(z).
\end{equation}
Fig.~\ref{fig:skyrmion} shows plots for $\tilde{\Omega}(w)$ corresponding to the transformation of the bimerons in Fig.~\ref{fig:bimeron}.
The resulting configurations are identified as skyrmions.
In particular, the M\"obius transformation maps the antivortex to the center of the skyrmion, whereas the vortex occupies the rest of the plane extending to spatial infinity.
Approximately, the antivortex is mapped inside the circle $|w|=1$ and the vortex outside it.

In Fig.~\ref{fig:skyrmion}, we observe circular vortex contours in the far field (corresponding to the vortex) and circular to elongated contours in the skyrmion center (corresponding to the antivortex).
The dent in the shape of the contours below the skyrmion center is attributed to the region where $m_3>0$ below the antivortex, as noted in connection with Fig.~\ref{fig:bimeron} in Sec.~\ref{sec:bimeron}.
Finally, for smaller values of $\dm$ the M\"obius transformed bimeron solutions $\tilde{\Omega}(w)$ approach progressively an axially symmetric profile as expected for the BP solution \eqref{eq:skyrmion}.

\begin{figure}[t]
    \centering
    (a)
    \includegraphics[width=4.0cm]{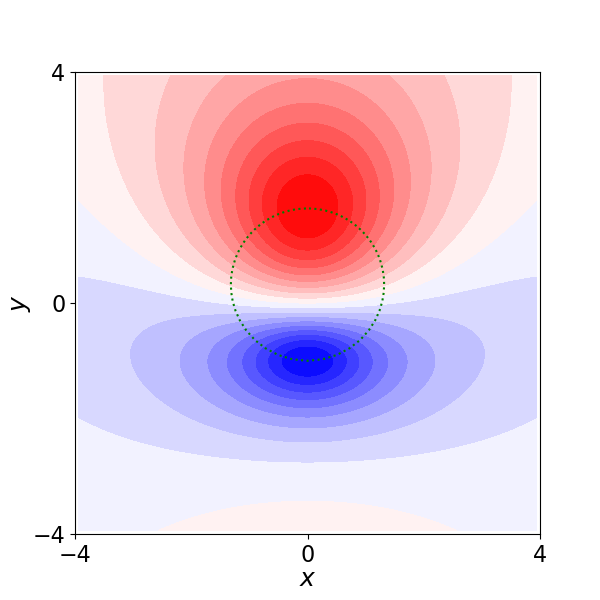}
    \includegraphics[width=4.0cm]{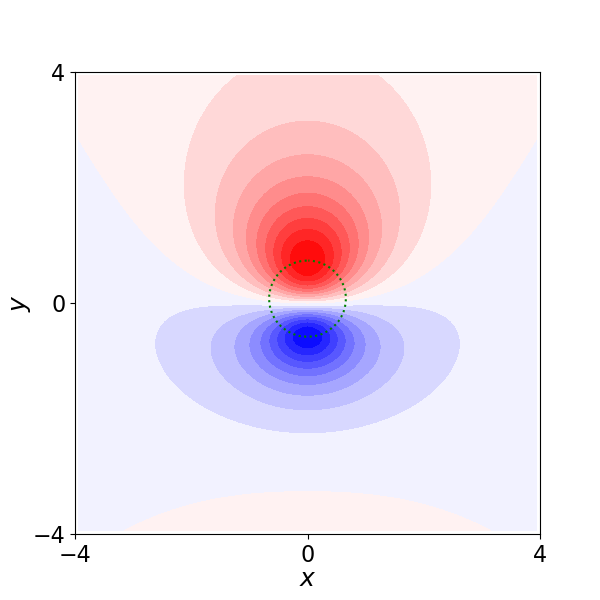}
    (b)
    \includegraphics[width=8.0cm]{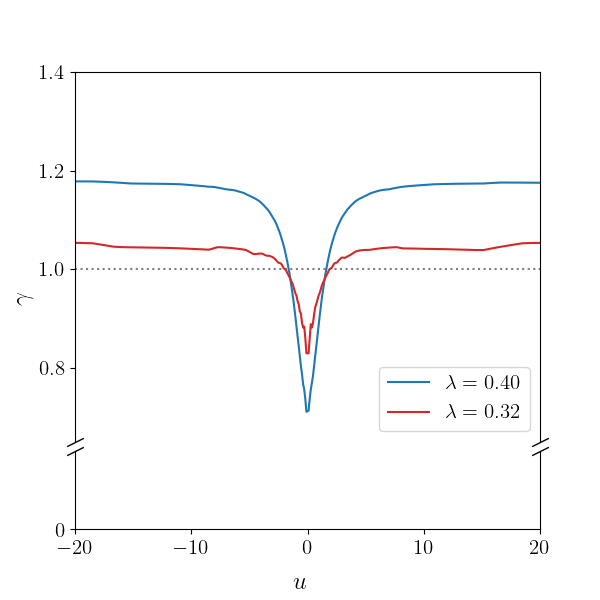}
    \caption{
    (a) The dotted line shows the path for $v=0$ on the $z$ plane for $\dm=0.40$ (left) and $\dm=0.32$ (right).
    (b) The quantity \eqref{eq:rOmega} calculated along the $u$ axis (for $v=0$) for $\dm=0.40$ and $\dm=0.32$.
    The value of $\gamma$ obtained at $u=0$ indicates the antivortex radius $R_2=\gamma R < R$ and the value at $|u|\to\infty$ indicates the vortex radius $R_1=R/\gamma < R$.
    They are both found to be smaller than the radius $R$, in Eq.~\eqref{eq:vortexRadius} of the exchange model.
    }
    \label{fig:linePlots}
\end{figure}

We expect the radii of the two merons to be different, and, in fact, the radius of the antivortex $R_2$ to be smaller than the radius $R_1$ of the vortex.
Following the discussion in Ref.~\cite{1978_NPB_Gross}, we expect a M\"obius transformed configuration $\tilde{\Omega} = \gamma_2/(iw)$ around the center of the skyrmion and $\tilde{\Omega} = \gamma_1/(iw)$ in the outer region (for $|w| \gg 1$), with $\gamma_2 < \gamma_1$.
This would imply vortex and antivortex radii $R_1=R/\gamma_1$ and $R_2=\gamma_2 R$ respectively, with $R$ defined in \eqref{eq:vortexRadius}.
In order to detect this behavior in the chiral skyrmion configurations, we consider the relative scaling
\begin{equation} \label{eq:rOmega}
\gamma := |w \tilde{\Omega}|
\end{equation}
which is expected to yield $\gamma_1, \gamma_2$ for the corresponding regions.

Figure~\ref{fig:linePlots}b shows the scaling parameter $\gamma$ as obtained along the horizontal $u$-axis (i.e. the line $v=0$) in Fig.~\ref{fig:skyrmion} for the two values of $\dm$.
The saturation of the scaling parameter to $\gamma_1>1$ for $|u|\gg 1$ indicates a vortex radius $R_1 < R$.
Similarly, for the antivortex, we find $\gamma_2<1$ for $u\ll1$, corresponding to $R_2<R$.
For $\dm=0.32$, the values of $\gamma_1, \gamma_2$ are not very far from unity, showing that, in this case, the bimeron is close to the Belavin-Polyakov solution in Eq.~\eqref{eq:BPmerons}.
On the other hand, for $\dm=0.40$, the values of $\gamma_1, \gamma_2$ deviate significantly from unity indicating that the individual meron sizes are shrinking compared to the total bimeron size and they are thus progressively detaching from each other.

The separation of merons is in direct correspondence to the description in Ref.~\cite{1978_NPB_Gross}.
Related are also experimental reports in an easy-plane biaxial magnet where ``bubble domains'' are robustly observed \cite{2017_ncomms_ChenLiu}.
Specifically, the so-termed ``cyan bubble domain'' contains two separated vortices or merons and it can be considered as the experimental realization of meron detachment.

\begin{figure}[t]
    \centering
    (a)\includegraphics[width=9.5cm]{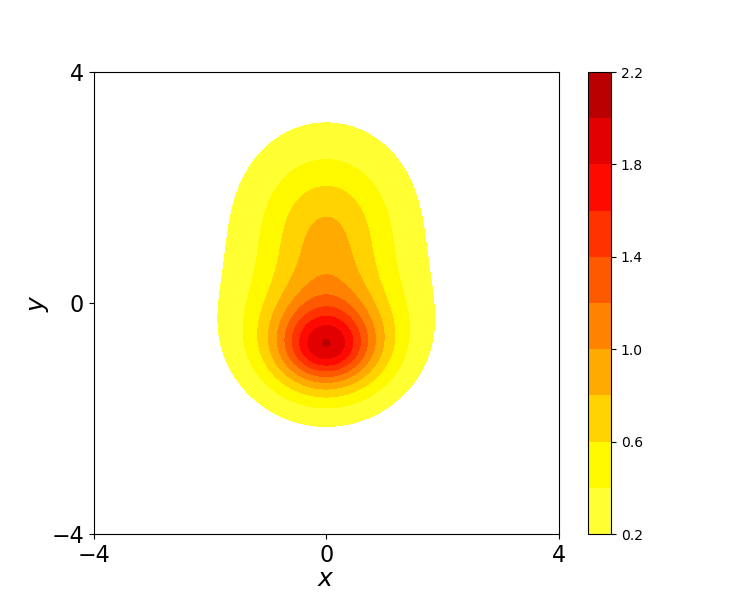}
    (b)\includegraphics[width=9.5cm]{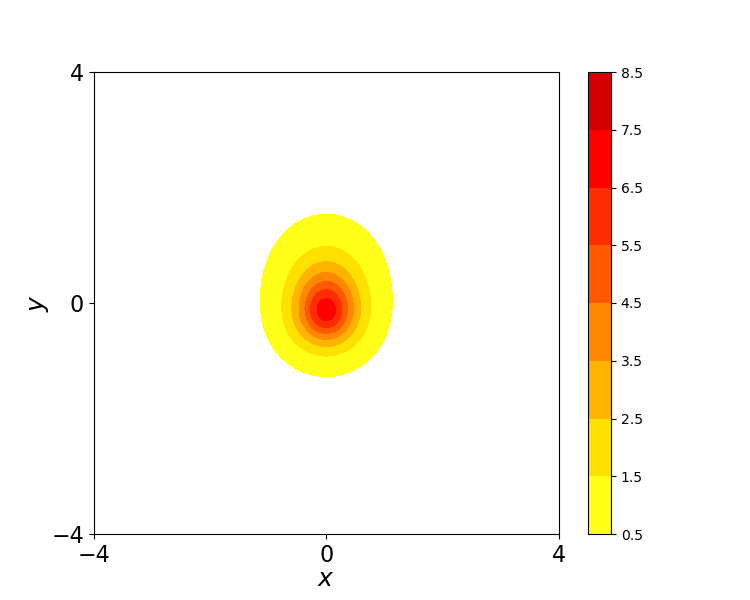}
    \caption{
    Contour plots for the topological density $\skyrmion$, defined in Eq.~\eqref{eq:skyrmionNumber}, of the bimeron solutions shown in Fig.~\ref{fig:bimeron} for  (a) $\dm=0.4$, (b) $\dm=0.32$.
    The same number of contours is plotted in both figures.
    }
    \label{fig:bimeron_topo}
\end{figure}

Let us now proceed to consider the topological density distribution of the bimeron solutions.
One should have in mind that the topological density for the rational map \eqref{eq:BPmerons} is axially symmetric (despite the asymmetry of the configuration).
Figure~\ref{fig:bimeron_topo} shows the topological density for the bimeron solutions of the two configurations in Fig.~\ref{fig:bimeron}.
For larger $\dm$, the topological density distribution has an elongated shape with the two merons contributing in different parts of space.
The topological density maximum is shifted to the antivortex side due to the sharper localization of the antivortex.
As $\dm$ decreases, the topological density distribution approaches axial symmetry, and its center is shifted closer to the center of the bimeron (at the origin).
These features further corroborate the assumption that the chiral bimeron approaches the configuration \eqref{eq:skyrmion} for $\dm\to 0$, while for large $\dm$ the two merons detach from each other.
A similar phenomenon was reported within a non-chiral model for anisotropic ferromagnets with competing interactions in \cite{2017_PRL_KharkovMostovoy}.

 \section{Bimeron pairs}
\label{sec:unbinding}

\begin{figure}[t]
    \centering
    \includegraphics[width=8.0cm]{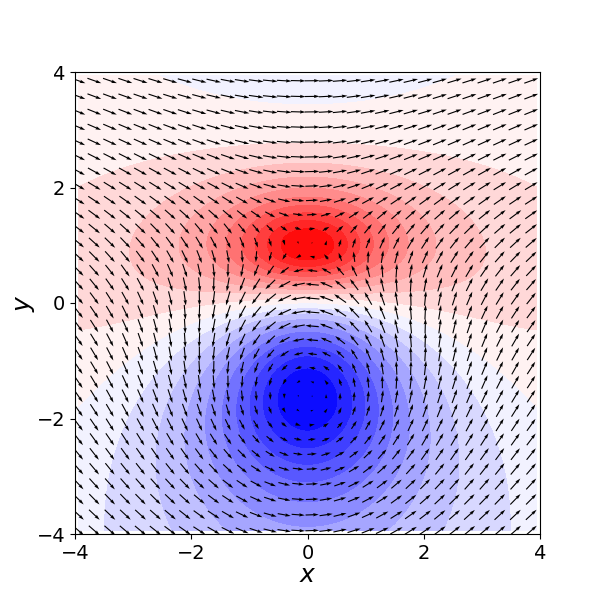}
    \caption{
    A bimeron solution of model \eqref{eq:LLG} for $\dm=0.4$ where the vortex points down and the antivortex points up.
    Contour plots for $m_3$ are colored, where red indicates $m_3 > 0$ and blue indicates $m_3 < 0$.
    The center of the bimeron, defined to be at the point where $m_1=-1$, has been placed at the origin.
    The skyrmion number is $\Skyrmion=1$.
    }
    \label{fig:bimeron-down}
\end{figure}

Considering the same system as in the previous sections, that is, retaining the vacuum magnetization $\magn=\e_1$, a second bimeron configuration can be found with an opposite skyrmion number.
This has been noted in Ref.~\cite{2019_PRAppl_MoonHwang}.
The second bimeron may be achieved by applying a rotation in space by $\pi$ ($x\to-x, y\to-y$) and reversing the third magnetization component ($m_3\to-m_3$) of the initial solution, in Fig.~\ref{fig:bimeron}.
The result, for $\dm=0.4$, is shown in Fig.~\ref{fig:bimeron-down}.
The vortex has now negative polarity and the antivortex has positive polarity.
This leads to a skyrmion number $\Skyrmion=-1$, opposite to the skyrmion number of the previously presented bimerons in Fig.~\ref{fig:bimeron}.
Thus, we conclude that the easy-plane ferromagnet can support chiral bimerons with opposite skyrmion numbers.

\begin{figure*}[t]
    \centering
    (a) \includegraphics[width=0.4\textwidth]{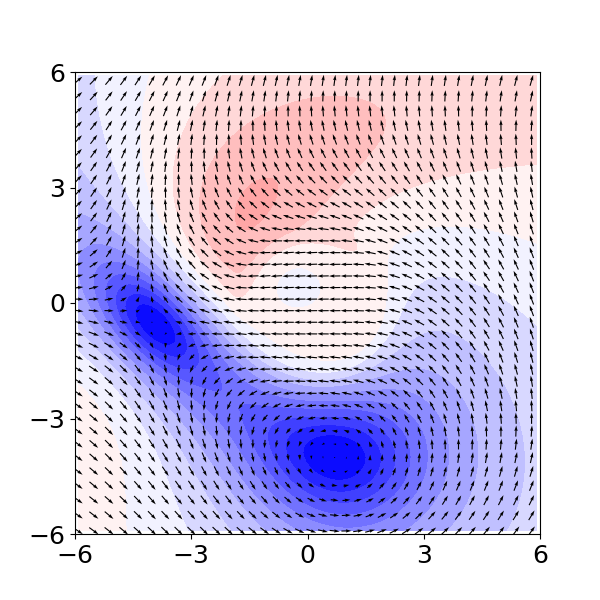}
    (b) \includegraphics[width=0.4\textwidth]{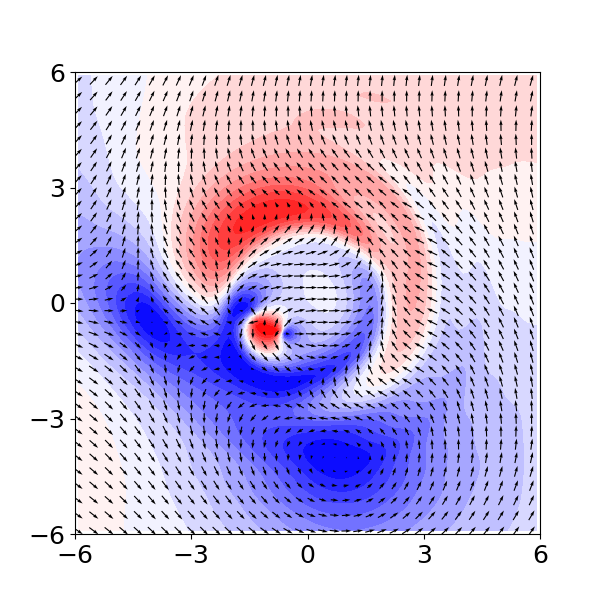}
    (c) \includegraphics[width=0.4\textwidth]{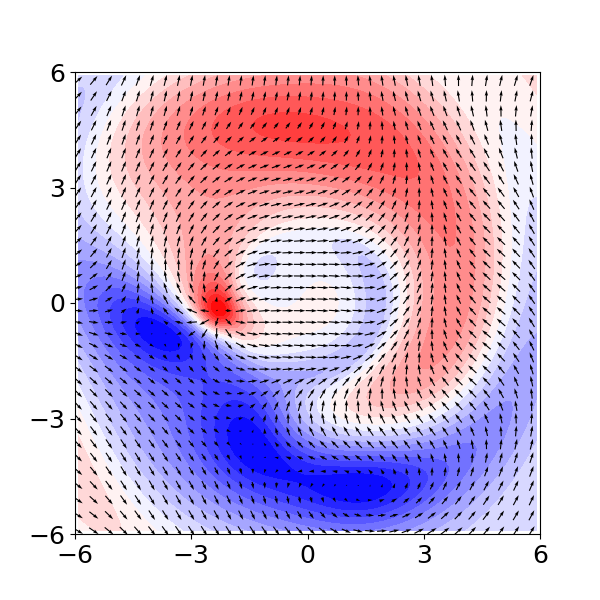}
    (d) \includegraphics[width=0.4\textwidth]{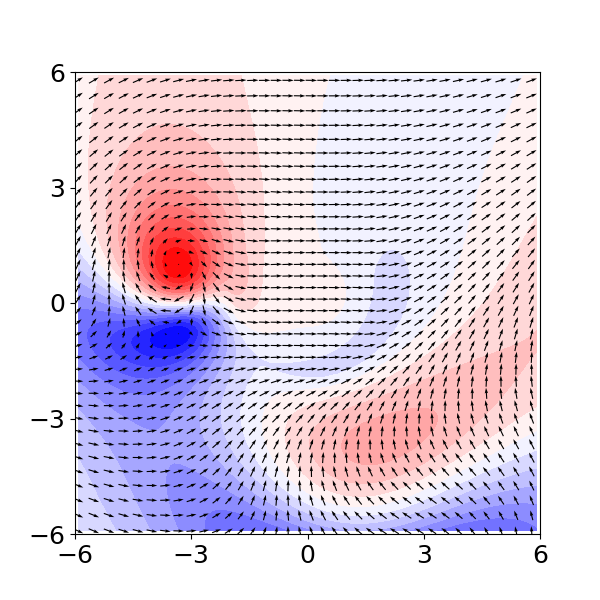}
    \caption{Spin torque is applied on an almost polarized state.
    The following snapshots are shown.
    (a) A vortex-antivortex is generated, after the application of spin torque with $\storque=-8$ in a circular region with radius 2 for $20$ time units.
    Following this stage, the polarization is reversed by setting $\storque=8$, and we show snapshots at
    (b) $t=20.6$, where a second vortex-antivortex pair with polarity down and a third one with polarity up are created, (c) $t=22.0$, where a bimeron has been formed, and (d) $t=28.0$, where only a single bimeron has remained in the system.
    }
    \label{fig:simulation}
\end{figure*}

A mechanism for the generation of bimerons can be readily suggested based on the existence of the two oppositely charged bimerons.
One may imagine the generation of a vortex-antivortex pair where both vortices have the same polarity, say down, and the simultaneous generation of a vortex-antivortex pair with polarity up \cite{1999_NL_PapanicolaouSpathis}.
Both pairs are topologically trivial, with $\Skyrmion=0$, and they can thus be created in a smooth way.
An exchange of partners between the two pairs would give recombination of the vortices and antivortices such that two bimerons with opposite skyrmion numbers would emerge.
When these unbind, they give the two bimerons in Fig.~\ref{fig:bimeron} and Fig.~\ref{fig:bimeron-down}.
A simulation demonstrating a pair of vortices together with a pair of antivortices, that may be interpreted as a pair of bimerons, was reported in \cite{2017_PRB_Meynell}.

We argue that the described mechanism may well be favored by the physics of the system because it has the following advantages.
A topologically trivial vortex-antivortex pair can be created out of fluctuations of the polarized state.
This pair is, though, a propagating structure \cite{1999_NL_PapanicolaouSpathis} and it would eventually be annihilated via energy dissipation.
On the other hand, a bimeron, once formed, is a static structure, that is, it is an energy minimum and thus a stable topological configuration that is robust against damping and perturbations.

A proof-of-concept numerical simulation that realizes the procedure has been performed.
We consider an initially polarized state along the $\magn=\e_1$ direction, but we add a perturbation by setting $\magn\approx (0.98,0.2,0)$ in the region $-2 \leq x,y \leq 1$.
Spin-transfer torque of the Slonczewski type is then applied with polarization along $x$.
The dynamics is described by 
\begin{equation} \label{eq:LLGS}
\p_\tau\magn = -\magn \times \heff + \alpha \magn\times\p_\tau\magn - \storque \magn\times(\magn\times\e_1).
\end{equation}
We set $\storque=-8$ and the polarized current is only applied in a circular region of radius 2 around the origin.
The negative value of $\storque$ corresponds to reversing the polarization to $-\e_1$ or to reversing the direction of the current.
We use the parameter value $\dm=0.4$ and damping $\alpha=0.1$.
A vortex-antivortex pair with negative polarity is created due to the dynamics and it is shown in Fig.~\ref{fig:simulation}a at time $t=20$.

Immediately after $t=20$, we reverse the spin torque parameter to $\storque=8$.
We see that a second vortex-antivortex pair, again with polarity down, is created, and soon after that, a third vortex-antivortex pair with polarity up starts growing.
Figure~\ref{fig:simulation}b shows the configuration at $t=20.6$.
Then, the antivortex from the third pair is annihilated with the vortex from the second pair and the remaining vortex of the third pair binds with the antivortex of the first pair, as shown in Fig.~\ref{fig:simulation}c, at time $t=22.0$.
The remaining vortex from the first pair and antivortex from the second pair have the same polarity, they approach each other, propagate away and eventually annihilate smoothly.
The picture shown in Fig.~\ref{fig:simulation}d shows the remaining bimeron at time $t=28.0$.

The mechanism for the generation of two bimerons can be generalized to a process where a collection of bimerons is generated while the total skyrmion number of the system remains zero.
If the temperature is taken into account, an easy-plane chiral magnet could give a gas of bimerons, leading to the question of a topological phase transition in the system.
Given that the bimerons are static states, these will not only be sustained due to thermal fluctuations but the system may be trapped in a state of multiple bimerons that may be a local energy minimum.

\section{Concluding remarks}
\label{sec:conclusions}

We have studied in detail chiral bimeron solutions in a magnet with Dzyaloshinskii-Moriya interaction.
We found that they bear essential similarities to the bimerons originally discussed within the O(3) nonlinear sigma model and we have quantified the similarity of the features employing a M\"obius transformation of the bimeron configuration.
The chiral bimerons thus appear to be a realization of the original bimeron configurations, presenting, for example, the possibility to detach from each other.
This opens the possibility of the proliferation of merons under temperature or due to an external probe.
We have described a method and have given a proof-of-concept simulation for the generation of bimerons.

We have identified the range of parameters for the existence of bimerons and the configuration dependence on the parameter values.
We further identified a special feature in the configuration, that is, the change of the sign of the out-of-plane magnetization component $m_3$ in the area beyond the antivortex.

A remarkable property of chiral bimerons is that they are static solutions of the model, unlike the situation in non-chiral magnets where vortex pairs are necessarily dynamical and they rotate around each other.
The dynamics of chiral bimerons thus emerges as an interesting topic to study.

\section*{Acknowledgements}
We are grateful to Riccardo Tomasello for discussions on many of the issues discussed in this work.
This work was supported by the project “ThunderSKY” funded by the Hellenic Foundation for Research and Innovation and the General Secretariat for Research and Innovation, under Grant No. 871.

\bigskip


%

\end{document}